\documentclass[runningheads]{llncs}
\usepackage[T1]{fontenc}
\usepackage{graphicx}
\usepackage{graphicx}
\usepackage{amsmath}
\usepackage{multirow}
\usepackage{booktabs}
\usepackage{array}
\usepackage{subfigure}
\usepackage{amssymb}
\usepackage{algorithm}
\usepackage{verbatim}
\usepackage{bm}     
\usepackage{enumitem}       

\usepackage{algpseudocode}

\begin{document}
\title{Multi-granularity Item-based Contrastive Recommendation}
%
%
\author{Ruobing Xie\thanks{Ruobing Xie and Zhijie Qiu have equal contributions.} \and Zhijie Qiu \and Bo Zhang \and Leyu Lin}
\institute{WeChat, Tencent \\
}

\maketitle              
\begin{abstract}
Contrastive learning (CL) has shown its power in recommendation. However, most CL-based recommendation models build their CL tasks merely focusing on the user's aspects, ignoring the rich diverse information in items. In this work, we propose a novel Multi-granularity item-based contrastive learning (MicRec) framework for the matching stage (i.e., candidate generation) in recommendation, which systematically introduces multi-aspect item-related information to representation learning with CL. Specifically, we build three item-based CL tasks as a set of plug-and-play auxiliary objectives to capture item correlations in feature, semantic and session levels. The feature-level item CL aims to learn the fine-grained feature-level item correlations via items and their augmentations. The semantic-level item CL focuses on the coarse-grained semantic correlations between semantically related items. The session-level item CL highlights the global behavioral correlations of items from users' sequential behaviors in all sessions. In experiments, we conduct both offline and online evaluations on real-world datasets, verifying the effectiveness and universality of three proposed CL tasks. Currently, MicRec has been deployed on a real-world recommender system, affecting millions of users. The source code will be released in the future.
\keywords{Recommendation  \and Contrastive learning \and Matching.}
\end{abstract}

\section{Introduction}

Real-world recommender systems aim to provide items according to user preferences from million-level item candidates \cite{guo2017deepfm}. It is challenging to directly conduct complicated ranking algorithms that involve user-item interactions on all items, for even the linear complexity with million-level corpus size is unacceptable \cite{zhu2018learning,xie2020internal}. Hence, practical recommender systems usually adopt the classical two-stage architecture \cite{covington2016deep} to balance effectiveness and efficiency, which consists of both ranking and matching modules. The \emph{matching} module \cite{xu2018deep} (i.e., candidate generation \cite{covington2016deep}) tries to retrieve a small subset of (usually hundreds of) item candidates from the million-level large corpora. In contrast, the \emph{ranking} module focuses on the specific ranks of top items retrieved by matching for the final display.

Matching should jointly consider recommendation accuracy, efficiency, and diversity. The two-tower architecture is the mainstream matching method widely used in large-scale systems \cite{yi2019sampling,xie2020internal}. Precisely, two neural networks are conducted to learn user and item representations separately. The similarities between user and item representations are viewed as click probabilities for online serving \cite{cen2020controllable}.
However, existing models mainly concentrate on the CTR-oriented objectives, struggling with the \emph{data sparsity} and \emph{popularity bias} issues \cite{yao2021self,zhou2021contrastive}. These issues are even more serious in matching, since matching should deal with million-level item candidates and the supervised user-item labels are more sparse relatively.
Recently, with the blooming of contrastive learning (CL) \cite{chen2020simple},  plenty of CL-based recommendation models are proposed, which alleviate the sparsity and popularity bias issues via more sufficient and diverse training from additional self-supervised learning (SSL) signals \cite{zhou2020s3}. However, most CL-based models merely depend on different \emph{user augmentations} from various aspects such as user behaviors \cite{wu2022multi,xia2021self,xie2022contrastive} and user attributes \cite{xiao2021uprec,yu2021socially}, ignoring the rich multi-granularity information in \emph{items}. Furthermore, the item representation often has a more direct influence on the matching results in the two-tower architecture, which multiplies the significance of learning good item representations via CL in matching.

\begin{figure*}[!hbtp]
\centering
\includegraphics[width=0.88\textwidth]{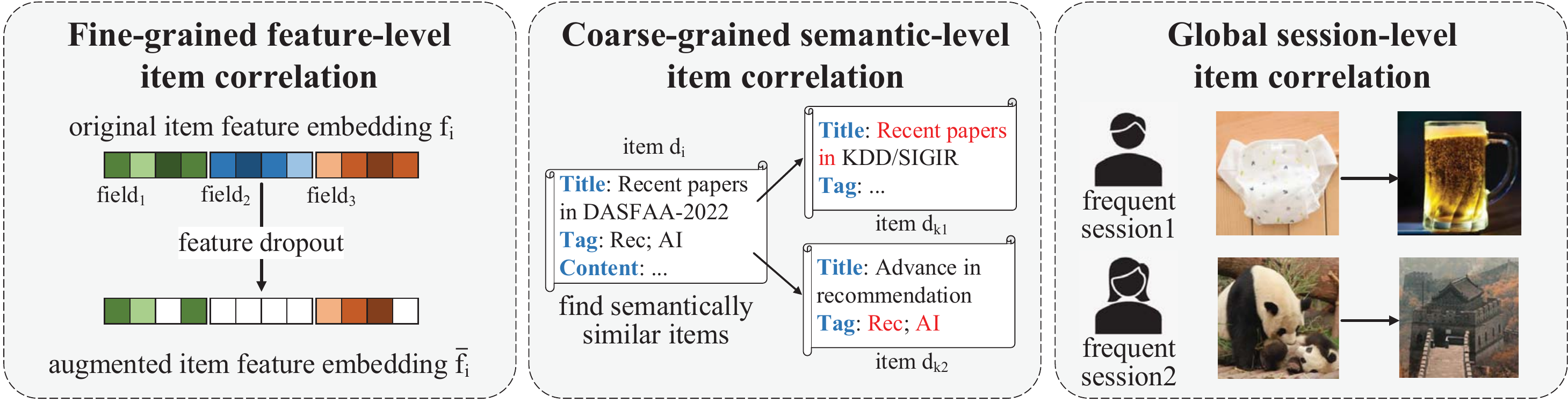}
\caption{Multi-granularity item correlations in real-world recommender systems.}
\label{fig:example}
\end{figure*}

In this work, we attempt to systematically conduct \textbf{item-based contrastive learning tasks} for matching. The key point is how to build more \emph{effective and diverse item augmentations}. We find that an item could be described from multiple granularities, such as its features, semantics, and behavioral information. Fig. \ref{fig:example} shows three different aspects of item correlations.
(a) The \emph{fine-grained feature correlations} of items reflect the feature-level similarities between items and their augmentations built via feature dropouts. (b) The \emph{coarse-grained semantic correlations} highlight items' semantic similarities extracted from titles, tags or contents. Differing from feature correlations, semantic correlations measure the semantic relevances between two items. (c) The \emph{global session-level correlations} capture items' behavioral similarities via users' sequential behaviors in all sessions. For example, the [diaper$\rightarrow$beer] and [panda$\rightarrow$the Great Wall] in Fig. \ref{fig:example} are frequent sessions that reveal surprising but reasonable item associations beyond similar features and semantics.
These three item correlations help to build a comprehensive understanding of all items from different aspects, which is essential especially in matching that should handle million-level (long-tail) items.

Therefore, we propose a new \textbf{Multi-granularity Item-based Contrastive Recommendation (MicRec)} framework, aiming to encode the under-explored item correlations into representation learning via CL tasks. Specifically, we design three item-based CL tasks.
(1) The \textbf{\emph{feature-level item CL}} focuses on fine-grained feature correlations. It builds item augmentations via feature element/field dropouts as positive samples, improving the alignment and uniformity of item representations that are beneficial as \cite{wang2020understanding}.
(2) The \textbf{\emph{semantic-level item CL}} attempts to model item similarities from the coarse-grained semantic aspect. Relevant items sharing the same item taxonomies (e.g., tags/categories) or similar titles/contents are viewed as the positive item pairs in this CL task.
(3) The \textbf{\emph{session-level item CL}} captures more implicit item behavioral similarities from the global view. Item pairs with high co-occurrences counted from all sessions are considered as positive samples.
We conclude the advantages of MicRec as follows:
(1) the multi-granularity item-based CL tasks work as auxiliary SSL losses, which bring in additional training signals to fight against sparsity, making both popular and long-tail items get more sufficient and diversified training besides the CTR loss.
(2) The three CL tasks encode more diverse item correlations (feature, semantics, and session levels) into item representations rather than simply relying on behaviors, which is also beneficial for diversity.
(3) MicRec is a plug-and-play framework that could be conveniently deployed on almost all matching models. Its universality and simplicity are welcomed by industrial systems.

In experiments, we evaluate MicRec on large-scale recommendation datasets and achieve significant improvements over competitive baselines. We also verify its effectiveness in an online evaluation. Moreover, we conduct an ablation study and a universality analysis of MicRec. The contributions of this work are as:
\begin{itemize}
 \item We highlight the importance of item-based CL in recommendation beyond lots of widely-explored user-based CL models. To the best of our knowledge, we are the first to systematically consider multi-granularity item-based CL.
 \item We propose a set of item-based CL tasks from feature, semantic, and session aspects, which could jointly improve accuracy and diversity. Our item-based CL tasks can be easily deployed on most matching models and scenarios.
 \item The significant improvements in offline and online evaluations verify the effectiveness and universality of MicRec. Currently, MicRec has been deployed on a real-world recommendation system, affecting millions of users.
\end{itemize}

\section{Related Work}

\textbf{\emph{\quad Matching in Recommendation}.}
The two-tower architecture is the mainstream method in practical matching, which learns user and item representations separately in two towers \cite{covington2016deep,yao2021self}. Different feature interaction models \cite{guo2017deepfm,song2019autoint} could be used in user/item towers for feature interaction. Some two-tower models \cite{sun2019bert4rec,xie2020internal} concentrate on user behavior modeling in user tower.
Some models \cite{cen2020controllable} focus on user's multi-interest modeling.
Other architectures such as
Tree-based methods \cite{zhu2018learning}, end-to-end deep retrieval \cite{gao2021learning}, and graph-based methods \cite{xie2021improving,zheng2021dgcn} are also explored in matching.
In this work, we attempt to achieve a deeper understanding of items. To verify the universality, we deploy MicRec on various matching models (e.g., YoutubeDNN \cite{covington2016deep}, AutoInt \cite{song2019autoint}, and ICAN \cite{xie2020internal}). It is also easy to deploy MicRec on models with sophisticated user modeling.

\textbf{\emph{Contrastive Learning in Recommendation.}}
Recently, SSL and CL have been verified in recommendation, which successfully address the sparsity and popularity bias issues \cite{zhou2021contrastive}.
Many of these models rely on user augmentations on user attributes and behaviors \cite{xie2020contrastive,xiao2021uprec}.
$\mathrm{S}^3$-Rec \cite{zhou2020s3} builds four CL tasks among items, attributes, sentences, and sub-sentences in sequential recommendation.
\cite{zhou2021contrastive} adopts disentangled CL-based model to candidate generation with multi-intention queues.
Other CL-based model focuses on graph learning \cite{xia2021self,yu2022graph}, cross-domain recommendation \cite{xie2022contrastive}, multi-behavior recommendation \cite{wu2022multi}, and social recommendation \cite{yu2021socially}.
\cite{yao2021self} is the most related work. It proposes two random and correlated feature masking (RFM/CFM) strategies, building augmentations with feature masking and dropout.
Differing from existing works that mainly focus on user-based augmentations or simply consider one aspect for item augmentation, we design a novel set of item-based CL tasks learning from different aspects of items. To the best of our knowledge, we are the first to jointly consider feature-level, semantic-level, and session-level item-based CL in recommendation.

\section{Methodology}
\label{sec.methodology}

\subsection{Preliminaries}
\label{sec.preliminaries}

Matching is the first stage of a classical two-stage recommendation framework \cite{covington2016deep}, which aims to efficiently retrieve hundreds of good items from million-level item corpora. In this work, we propose MicRec mainly for matching.

\textbf{Two-tower Architecture.}
The two-tower architecture is the most widely-used matching model in practice \cite{covington2016deep,xie2020internal}. A typical two-tower model contains two neural networks (i.e., towers) to construct user and item representations separately. Formally, we build the user and item representations $\bm{u}_i$ and $\bm{d}_k$ as $\bm{u}_i = \mathrm{DNN}^u(\bm{f}^u_i)$ and $\bm{d}_k = \mathrm{DNN}^d(\bm{f}^d_k)$, where $\mathrm{DNN}^u(\cdot)$ and $\mathrm{DNN}^d(\cdot)$ are neural networks and $\bm{f}^u_i$ and $\bm{f}^d_k$ are the input raw feature embeddings of user and item. The (cosine) similarity between $\bm{u}_i$ and $\bm{d}_k$ is viewed as the click probability of $(u_i,d_k)$. The classical sampled softmax is often used as the training objective \cite{cen2020controllable}.
In online serving, an approximate nearest neighbor server (e.g., \cite{johnson2019billion}) is used to fast retrieve items via user-item similarities. In the two-tower architecture, item representations have a direct impact on the final matching scores. In this case, the proposed item-based CL is more significant for item-focused learning.

\textbf{Multi-granularity Item Information.}
In practice, models are expected to make full use of all item features for more accurate and diverse recommendations. Here, we introduce three typical types of item information to be used in MicRec:
\begin{itemize}[leftmargin=*]
 \item \textbf{Item features}. We follow classical models \cite{song2019autoint,yao2021self} that divide item features into feature fields. Each field (e.g., ID, tag) is allocated with an embedding. An item and its feature-based augmentation have similar feature embeddings, which reflects the \emph{fine-grained feature correlations} with various features.
 \item \textbf{Item semantics}. Semantics is more coarse-grained information compared to feature fields. Semantic similarities are often reflected by similar item titles, contents, or taxonomies. This information helps to capture the \emph{coarse-grained semantic correlations} between two semantically related items.
 \item \textbf{Sessions}. Items sequentially clicked by a user in a session usually imply implicit and reasonable behavioral patterns (e.g., [diaper$\rightarrow$beer]). We calculate the global co-occurrences of items in the same session from the overall dataset. This information reflects the \emph{global session correlations} between two items.
\end{itemize}
All three types of item information widely exist and can be easily collected in real-world systems. These item correlations are used as additional SSL signals.

\subsection{Overall Framework}
\label{sec.overall_framework}

\begin{figure}[!hbtp]
\centering
\includegraphics[width=0.70\columnwidth]{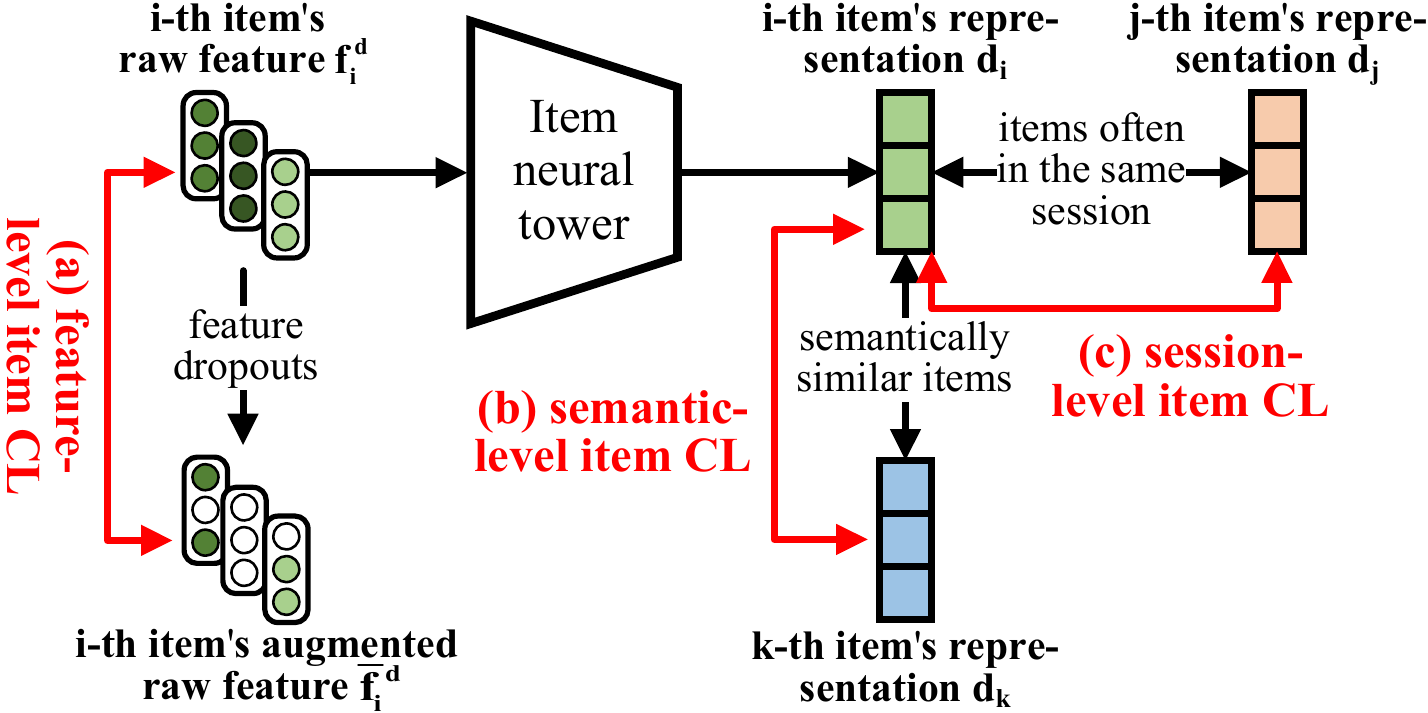}
\caption{The proposed three plug-and-play item-based CL tasks in MicRec. }
\label{fig:architecture}
\end{figure}

MicRec adopts the classical two-tower architecture as the base model (i.e., ICAN \cite{xie2020internal}), and proposes three item-based CL tasks as in Fig. \ref{fig:architecture}:
(a) The feature-level item CL considers an item and its feature-based augmentations as positive instances. We design different dropout strategies at both feature field and element levels to get diverse augmentations.
(b) In semantic-level item CL, two distinct items having similar semantics (e.g., titles, tags) are regarded as positive pairs in CL. It focuses on the semantic similarities between two items' representations.
(c) The session-level item CL provides a new train of thought for modeling item similarities via the global item co-occurrences in all sessions. It makes full use of the under-explored user global behaviors as an essential supplement to item learning.
Note that MicRec is effective and universal, which is convenient to be deployed on different datasets with most embedding-based matching models.

\subsection{Feature-level Item Contrastive Learning}
\label{sec.feature_CL}

In MicRec, the $i$-th input item embedding $\bm{f}_i^d$ is the concatenation of the item's field embeddings (e.g., trainable ID embedding $\bm{f}_i^I$, tag embedding $\bm{f}_i^T$, provider embedding $\bm{f}_i^P$). For each item embedding $\bm{f}^d_i$, we conduct a feature-level dropout to get the item's augmented embedding $\bm{\bar{f}}^d_i=\mathrm{Dropout}(\bm{f}^d_i)$. We define $g_f(\bm{f}^d_i,\bm{\bar{f}}^d_j)$ as the similarity score between the original $d_i$ and augmented $d_j$, noted as:
\begin{equation}
\begin{split}
g_f(\bm{f}^d_i,\bm{\bar{f}}^d_j) = ( \mathrm{MLP}_f(\bm{f}^d_i) )^\top \mathrm{MLP}_f(\bm{\bar{f}}^d_j).
\end{split}
\label{eq.g_geature}
\end{equation}
$\mathrm{MLP}_f (\cdot)$ is a multi-layer perception as a projector. Here, we want items to be closer to their augmentations than other items'. Therefore, the feature-level item CL loss $L_{fea}$ is defined following the classical InfoNCE \cite{chen2020simple} as follows:
\begin{equation}
\begin{split}
L_{fea} = - \sum_{d_i} \log \frac{\exp(g_f(\bm{f}^d_i,\bm{\bar{f}}^d_i)/\tau)}
{\sum_{d_j \in N_i} \exp (g_f(\bm{f}^d_i,\bm{\bar{f}}^d_j)/\tau)}.
\end{split}
\label{eq.L_feature}
\end{equation}
$d_j \in N_{i}$ are negative samples that $d_j \neq d_i$. $\tau$ is the temperature of CL. Different from \cite{chen2020simple} that conducts in-batch negative sampling, we directly utilize randomly sampled $d_j \neq d_i$ as negative samples to avoid excessive suppression of popular items, which is widely verified in our online matching modules. We provide three feature dropout strategies to build item augmentations for different demands:
\begin{itemize}[leftmargin=*]
  \item Element-level dropout, which randomly masks elements in any field of the raw feature embedding $\bm{f}^d_i$ with a certain mask ratio. In this case, any field embedding may be partially masked (e.g., $\bm{\bar{f}}^d_i=\mathrm{concat}(\bm{\bar{f}}_i^I, \bm{\bar{f}}_i^T, \bm{\bar{f}}_i^P)$).
  \item Field-level dropout, which randomly masks the whole feature fields (e.g., $\bm{\bar{f}}^d_i=\mathrm{concat}(\bm{f}_i^I, \bm{0}, \bm{f}_i^P)$ after the field-level dropout with the whole field $\bm{f}_i^T$).
  \item Categorial feature dropout, which is designed for categorial features having multiple values (e.g., tags). Some tags are randomly ignored when building the overall tag field embedding $\bm{f}_i^T$.
\end{itemize}
In experiments, we find that the simple combination of the field-level dropout and the categorial feature dropout performs the best, with the mask ratio set as $0.5$ to ensure a challenging and informative SSL training.

\subsection{Semantic-level Item Contrastive Learning}
\label{sec.semantic_CL}

The feature-level item CL focuses on the fine-grained raw feature correlations between items and their dropout-based augmentations.
Different from the feature-level CL, the semantic-level item CL suggests that an item should be similar to its semantically similar items (viewed as natural semantics-based augmentations of this item). For each item $d_i$, we first discover its semantically similar item set $P_i^t$ as the positive sample set in CL. We also randomly select other items not in $P_i^t$ to form the negative set $N_i^t$.
The similarity score $g_t(\bm{d}_i,\bm{d}_j)$ between $d_i$ and $d_j$ is calculated as $g_t(\bm{d}_i,\bm{d}_j) = ( \mathrm{MLP}_t(\bm{d}_i) )^\top \mathrm{MLP}_t(\bm{d}_j)$, where $\bm{d}_i = \mathrm{DNN}^d(\bm{f}^d_i)$ is the item representation. Formally, the semantic-level item CL loss $L_{sem}$ is:
\begin{equation}
\begin{split}
L_{sem} = - \sum_{d_i} \sum_{d_k \in P_i^t} \log \frac{\exp(g_t(\bm{d}_i,\bm{d}_k)/\tau)}
{\sum_{d_j \in N_i^t} \exp (g_t(\bm{d}_i,\bm{d}_j)/\tau)},
\end{split}
\label{eq.L_semantic}
\end{equation}
$L_{sem}$ takes advantage of the intrinsic coarse-grained semantic correlations of items besides sparse user-item behaviors, highlighting item semantics (one of the central item information) to improve item understanding. It also brings in more training opportunities for long-tail items with fewer interactions.

Item semantics can be represented by different attributes, such as item titles, taxonomies (e.g., categories, tags), and contents. We propose a series of metrics to collect the semantically similar items from different aspects as:
\begin{itemize}[leftmargin=*]
  \item For item titles, we adopt their sentence embeddings learned from a pre-trained language model (e.g., we use BERT \cite{devlin2019bert} in this work) as the title embeddings. The top-k nearest items retrieved by the cosine similarities between title embeddings are viewed as the semantically similar positive pairs in this CL.
  \item For taxonomies, we regard items having the same taxonomies as positive item pairs of each other, which is more generalized than similar item titles.
  \item For item contents such as videos and texts, we could also use pre-trained visual or textual encoders to learn the content embeddings as item semantics. The positive item pairs are selected via their content embeddings' similarities.
\end{itemize}
These attributes have their own pros and cons in representing semantics. Item titles are the major information exposed to users and have large impacts on CTR, while they may suffer from clickbaits. Item taxonomies are keywords of contents, while they are more generalized and may bring in noises. Item contents are more comprehensive, while their qualities strongly rely on the content encoders.
In our system, we find that considering title similarities in semantic-level CL achieves the best performance, since titles are sufficient to capture the main semantics for CTR, and clickbaits are largely pre-filtered.
Note that the semantic-level CL has the following major differences with the feature-level CL: (a) the semantic-level CL builds positive pairs via two items $(\bm{d}_i, \bm{d}_k)$ rather than $(\bm{f}^d_i,\bm{\bar{f}}^d_i)$. (b) The semantic-level CL only focuses on the semantic correlations, while the feature-level CL models the multi-factor feature correlations via feature dropout.

\subsection{Session-level Item Contrastive Learning}
\label{sec.session_CL}

Besides features and semantics, users' sequential behaviors in sessions also reflect item correlations. Items appearing in a session are very likely to have implicit correlations. Taking the classical [diaper$\rightarrow$beer] for example, these two items do have statistical behavioral correlations, which cannot be explicitly captured by feature or semantic similarities. Such global session correlations are good supplements to user-item behaviors and semantics for both accuracy and diversity.

To make full use of the sequential behavioral correlations in sessions, we propose a novel session-level item CL. Specifically, given a session $s=\{d_1, \cdots, d_n\}$ of $n$ clicked items, we let any two items $d_i, d_j \in s$ have one co-occurrence relation. Next, we scan all sessions in the train set to learn the co-occurrence matrix of all item pairs. For the $i$-th item, we build its positive sample set $P_i^s$ in the session view via a weighted sampling over its top-k co-occurred items (proportional to the co-occurrence counts). The $i$-th item's negative sample set $N_i^s$ are randomly selected from all items that have not co-occurred with $d_i$. We build $L_{sess}$ with a similar session-level similarity $g_s(\bm{d}_i,\bm{d}_j) = ( \mathrm{MLP}_s(\bm{d}_i) )^\top \mathrm{MLP}_s(\bm{d}_j)$, noted as:
\begin{equation}
\begin{split}
L_{sess} = - \sum_{d_i} \sum_{d'_k \in P_i^s} \log \frac{\exp(g_s(\bm{d}_i,\bm{d}'_k)/\tau)}
{\sum_{d_j \in N_i^s} \exp (g_s(\bm{d}_i,\bm{d}_j)/\tau)},
\end{split}
\label{eq.L_session}
\end{equation}
$L_{sess}$ assumes that items appearing together more times in a session should share the same user preferences and be more similar. It is \textbf{the first} attempt to encode the global session-level item correlations into item representations via CL.

\subsection{Specifications of MicRec}
\label{sec.specifications}

In MicRec, all three item-based CL tasks cooperate well with each other as auxiliary losses. We give detailed implementations of MicRec in our system.

MicRec is a universal and easy-to-deploy framework that can work with most embedding-based models. We follow the state-of-the-art multi-channel matching model ICAN \cite{xie2020internal} to build our MicRec's two-tower neural network, which has been verified in our online matching system. ICAN designs a domain-specific attention and a field-level self-attention to learn user behavioral features, and build user and item towers via three-layer MLPs. We inherit the main neural networks and user features of ICAN and remove its multi-domain designs. Formally, the final user representations $\bm{u}_i$ used for matching prediction is as follows:
\begin{equation}
\begin{split}
\bm{u}_i = \mathrm{MLP}_u(\mathrm{Concat}(\mathrm{Flatten} (\mathrm{Transformer}(\bm{F}_i^b)), \bm{u}_i^{o})),
\end{split}
\label{eq.user_representation}
\end{equation}
$\bm{F}_i^b$ is the raw behavior matrix of $u_i$. After a Transformer-based interaction, the user behavioral feature is combined with other user features $\bm{u}_i^{o}$ and fed into a 3-layer $\mathrm{MLP}_u$ to get $\bm{u}_i$. We also conduct another 3-layer MLP to model item feature interactions and get $\bm{d}_k$ for the k-th item.
We follow the classical sampled softmax matching loss $L_{S}$ \cite{xie2020internal} as our objective. $S^+$ represents the positive set (i.e., $u_i$ has clicked $d_k$), and $S^-$ is the randomly-sampled negative set.
\begin{equation}
\begin{split}
L_{S} = -\sum_{(u_i,d_k) \in S^+} \log \frac{\exp (\bm{d}_k^\top \bm{u}_i)}{\sum_{(u_i,d_j) \in S^-} \exp (\bm{d}_j^\top \bm{u}_i)}.
\label{eq:L_S}
\end{split}
\end{equation}
Following classical CL-based recommendations \cite{zhou2021contrastive}, we jointly learn the matching loss $L_{S}$ and three item-based CL losses $L_{fea}$, $L_{sem}$, and $L_{sess}$ in a multi-task learning manner with hyper-parameters $\lambda_1$, $\lambda_2$, and $\lambda_3$ as loss weights. We have:
\begin{equation}
\begin{split}
L = L_{S} + \lambda_1 L_{fea} + \lambda_2 L_{sem} + \lambda_3 L_{sess}.
\label{eq:L_all}
\end{split}
\end{equation}
We implement MicRec on this two-tower architecture, since (1) it is widely-used, effective, universal, and easy to deploy. (2) Its item representations have direct impacts on the final matching scores as in Eq. (\ref{eq:L_S}), which could give full play to the advantages of item-based CL.
It is also convenient to deploy MicRec on other matching models as plug-and-play objectives (see the universality analysis in Sec. \ref{sec.universality}), or along with other user-based CL tasks for deeper user modeling.

\section{Online Deployment and Complexity}
\label{sec.online_deployment}

We have deployed MicRec on a real-world recommender system, which is widely used by millions of users. MicRec functions as one of multiple matching channels with other modules unchanged, which is a classical online matching deployment setting.
In offline routine training, all item representations are learned via MicRec and indexed for online serving. The online serving is the same as the base model. The learned user representation of Eq. (\ref{eq.user_representation}) is used to fast retrieve top-K ($K=500$) similar items via user-item embedding similarities for the following ranking module.
We train MicRec on Tesla P40 with TensorFlow. Compared with the base model, MicRec adopts three additional CL losses, which inevitably brings in additional computation costs in offline training for each epoch. Nevertheless, the offline training time of MicRec is less than twice that of the base model, which is acceptable for daily training in our system. Note that MicRec \textbf{has equal online memory/computation costs} compared to its corresponding base model, since they have exactly the same online serving. Due to anonymity, we will give more detailed descriptions of our systems and deployment in the released version.

\section{Experiments}

\subsection{Experimental Settings for Matching}

\subsubsection{Datasets.}
\label{sec.dataset}

Since there are very few open datasets that have industrial item scales with practical feature and session information for matching, we build two large-scale datasets of different scenarios from a widely-used recommender system. The first Video-636M dataset contains nearly $25$ million users and $636$ million click instances. The second News-76M is a news recommendation dataset which has nearly $6.4$ million users and $76$ million click instances. These instances are then split into train and test sets in chronological order. We have $47$ million and $7.2$ million click instances as the test set in two datasets.
All data are preprocessed via data masking to protect user privacy. Table \ref{tab:dataset} shows the detailed statistics.

\begin{table}[!hbtp]
\centering
\small
\caption{Statistics of two large-scale industrial datasets for matching.}
\begin{tabular}{p{1.8cm}|p{1.8cm}<{\centering}p{1.8cm}<{\centering}p{1.8cm}<{\centering}p{1.8cm}<{\centering}}
\toprule
dataset & \# user & \# item & \# session & \# click \\
\midrule
Video-636M & 25,476,322 & 2,406,028 & 58,595,540 & 635,774,650 \\
News-76M & 6,364,403 & 1,056,630 & 11,455,920 & 76,303,835 \\
\bottomrule
\end{tabular}
\label{tab:dataset}
\end{table}

\subsubsection{Competitors.}

In this work, we implement several competitive matching models and CL-based recommendation models as our competitors.
\begin{itemize}[leftmargin=*]
 \item \textbf{Factorization Machine (FM).} FM \cite{rendle2010factorization} is a classical matching model widely used in industry. It uses latent vectors to capture second-order feature interactions. The user-item embedding similarity is used for fast retrieval.
 \item \textbf{YoutubeDNN.} YoutubeDNN \cite{covington2016deep} is also representative, which brings DNNs into the two-tower architecture to model high-order feature interactions.
 \item \textbf{DeepFM.} DeepFM \cite{guo2017deepfm}  combines DNN with neural FM models in parallel to build both user and item representations separately inside the user and item towers under the two-tower architecture.
 \item \textbf{AutoInt.} AutoInt \cite{song2019autoint} introduces self-attention to feature interaction modeling. For matching, we also adopt AutoInt separately inside two towers.
 \item \textbf{ICAN*.} ICAN \cite{xie2020internal} is one of the SOTA multi-domain matching models verified on industrial systems. For the single-domain matching, we follow ICAN's main neural network and remove its domain-specific designs (noted as ICAN*). Note that ICAN* is the same as the base model of MicRec in Sec. \ref{sec.specifications}.
 \item \textbf{RFM*.} \cite{yao2021self} is the SOTA item-based CL model. It proposes two feature-level item-based CL tasks, namely random feature masking (RFM) and correlated feature masking (CFM). CFM is not suitable for our datasets, since our items have very few categorial features that could have multiple values. Hence, we modify RFM with MicRec's feature dropout as RFM* (i.e., MicRec (fea)). For fair comparisons, RFM* is also deployed with the base network ICAN*.
 \item \textbf{MicRec's ablation versions.} We further conduct different combinations of CL tasks as different ablation versions. We regard \emph{fea} as the feature-level CL, \emph{sem} as the semantic-level CL, and \emph{sess} as the session-level CL.
\end{itemize}
We do not compare with existing customized CL-based ranking models \cite{zhou2020s3,xia2021self,xie2021adversarial,yu2021socially,zhou2021contrastive,wu2022multi} armed with user augmentations, since they focus on different tasks (e.g., ranking, disentangled recommendation, social recommendation) and could be deployed simultaneously with our item-based CL tasks.
MicRec is a universal framework that could be adopted on most embedding-based matching models (e.g., ICAN, YoutubeDNN, AutoInt, verified in Sec. \ref{sec.ablation}). Other neural structures for matching (e.g., \cite{cen2020controllable,xie2021improving}) and user-based CL tasks \cite{zhou2021contrastive,zhou2020s3} could also be easily deployed with MicRec, while it is not the core focus of this work.

\subsubsection{Parameter Settings.}
\label{sec.experimental_setting}

We deploy MicRec on ICAN. Following \cite{xie2020internal}, we use $20$ most recent click behaviors as input user behaviors. We have $8$ feature fields in Video-636M, and $17$ feature fields in News-76M following our online settings. The dimensions of item field embeddings (e.g., ID, tag), user field embeddings (e.g., gender, age), and item/user representations are set as $64$. The dimensions of the output embeddings in the 3-layer MLP are $128$, $64$, and $64$. All dimension sizes follow the base model ICAN \cite{xie2020internal} considering the online efficiency and effectiveness. For fair comparisons, all MicRec versions and baselines share the same user and item raw features (including semantic features), feature dimensions, and supervised objective of Eq. (\ref{eq:L_S}). In training, we use Adam with the learning rate $\alpha=0.001$, and set the batch size as $4,096$ according to the current online model.
Although a real-world model does have lots of hyper-parameters, we directly inherit most of them from the verified parameter settings of the current online base models.
We conduct a grid search for parameter selection on negative sample number among $\{10, 20, 50\}$, feature mask ratio among $\{0.3, 0.5, 0.7, 0.9\}$, and CL loss weights among $\{0.05, 0.1, 0.3, 0.5, 1.0\}$.
Empirically, we set the CL loss weights $\lambda_1=1.0$, $\lambda_2=0.3$, $\lambda_3=0.1$ for videos, and set $\lambda_1=1.0$, $\lambda_2=0.1$, $\lambda_3=0.1$ for news. We also find MicRec is robust with different temperatures and directly set $\tau=1$ for three CL tasks.
For the multi-granularity item-based CL, we conduct 1-layer projectors for all CL tasks, and select up to $50$ negative samples for efficiency. We consider two items within $1$ hour as an item co-occurrence in a session.
MicRec achieves good results without sophisticated CL designs and tricky parameter selections.
We conduct $5$ runs to get the results of models.

\subsection{Offline Evaluation on Matching (RQ1)}
\label{sec.offline_evaluation}

\subsubsection{Evaluation Protocols.}

\textbf{\emph{Matching should only value whether good items are retrieved, not the specific item ranks}}, since the top-N items retrieved by matching will be re-ranked by the following ranking modules. Hence, following classical matching models \cite{xie2020internal,zhou2021contrastive,xie2022contrastive}, we only utilize the top $N$ hit rate (HIT@N) as our matching metric instead of ranking metrics such as NDCG and AUC (since they are related to item ranks not considered in matching). We report HIT@N with different larger N \cite{gao2021learning,xie2020internal,xie2022contrastive} such as $50$, $100$, $200$, and $500$ to simulate the real-world matching modules, which often select hundreds of items from million-level candidates in practice (e.g., we retrieve top-$500$ items in online matching).

\begin{table*}[!hbtp]
\centering
\small
\caption{Offline evaluation. All improvements are significant (t-test with $p<0.01$).}
\begin{tabular}{l|c|c|c|c|c|c|c|c}
\toprule
\multirow{2}{*}{Model} & \multicolumn{4}{c|}{Video-636M} & \multicolumn{4}{c}{News-76M}\\
\cmidrule{2-9}
 & hit@50 & hit@100 & hit@200 & hit@500 & hit@50 & hit@100 & hit@200 & hit@500 \\
\midrule
FM & 0.1731 & 0.2487 & 0.3219 & 0.4495 & 0.2466 & 0.3072 & 0.3871 & 0.4843 \\
YoutubeDNN & 0.1779 & 0.2576 & 0.3375 & 0.4719 & 0.2691 & 0.3307 & 0.4116 & 0.5188 \\
DeepFM & 0.1825 & 0.2587 & 0.3368 & 0.4776 & 0.2659 & 0.3268 & 0.4083 & 0.5171 \\
AutoInt & 0.1859 & 0.2626 & 0.3552 & 0.4857 & 0.2716 & 0.3431 & 0.4213 & 0.5316 \\
ICAN* & 0.1872 & 0.2627 & 0.3563 & 0.4873 & 0.2748 & 0.3438 & 0.4241 & 0.5403 \\
RFM* & 0.1967 & {0.2752} & 0.3671 & 0.5065 & 0.2911 & 0.3609 & 0.4413 & 0.5605 \\
\midrule
MicRec & \textbf{0.2047} & \textbf{0.2831} & \textbf{0.3825} & \textbf{0.5209} & \textbf{0.3018} & \textbf{0.3719} & \textbf{0.4543} & \textbf{0.5758} \\
\bottomrule
\end{tabular}
\label{tab:offline}
\end{table*}

\noindent
\textbf{Experimental Results on Accuracy.}
From Table \ref{tab:offline} we can find that:

(1) MicRec achieves consistent improvements on all metrics compared with all baselines and ablation versions. The improvements over all baselines are significant with $p<0.01$ via t-test. The deviations of MicRec models on all metrics are less than $\pm0.002$. Note that the $2.8\%-4.1\%$ consistent improvements over the SOTA item-based CL model RFM* are impressive on the huge industrial datasets. It indicates that our multi-granularity item-based CL could well capture multi-granularity item correlations from feature, semantic, and session levels. These additional and diverse SSL signals brought by item-based CL help to learn better item representations and improve the \emph{alignment} and \emph{uniformity} stated in \cite{wang2020understanding}, and thus enable better performances in matching.

(2) The improvements of MicRec mainly derive from the three CL tasks. The sparsity issues in matching can be largely addressed via the multi-granularity SSL. Precisely, the feature-level CL improves the alignment and uniformity of items as classical CL models \cite{wang2020understanding,yao2021self}. The semantic-level CL highlights the semantic item correlations besides user-item interaction matrix, and semantic similarities have been verified in matching \cite{xie2021improving}. The session-level CL strengthens the sequential behavioral information to capture implicit but surprising item correlations that cannot be learned from $L_S$, $L_{fea}$, and $L_{sem}$ (e.g., \emph{diaper}$\rightarrow$\emph{beer}).
The three proposed CL losses are essential supplements to under-optimized items.
Moreover, considering various item correlations brings in more diversities compared with simply using user-click-item behaviors, which is also beneficial in matching.
Sec. \ref{sec.ablation} gives further ablation studies on three item-based CL tasks.

(3) Analyzing the results on two datasets, we know that MicRec has consistent improvements in both video and news domains, implying its universality. Moreover, MicRec has more significant improvements with larger N, which confirms its adaptability in real-world matching that usually retrieves hundreds of items. We also verify that MicRec can be deployed on different matching models (in Sec. \ref{sec.universality}). The improvements on different datasets and matching models confirm the universality of MicRec as an effective plug-and-play module.

\subsubsection{Experimental Results on Diversity.}

We also evaluate the diversity of MicRec via a classical aggregate diversity metric named \textbf{item coverage} \cite{herlocker2004evaluating,xie2021improving}. It represents the percentage of items in a dataset that the recommendation model is able to provide predictions for in the test set, which focuses on the overall diversity of matching. It is quite suitable to measure diversity via item coverage in matching, since matching is the upstream of ranking which should be responsible to make sure different items can be exposed to users. As in Table \ref{tab:ablation_test}, MicRec outperforms its base model ICAN on item coverage by a large margin from $12.4\%$ to $16.1\%$. The \textbf{$\bm{29.3\%}$ diversity improvement} shows that MicRec can alleviate popularity bias and improve diversity via the proposed set of item-based CL, which makes full use of different aspects of item information besides clicks. We further analyze the source of this diversity improvement in Sec. \ref{sec.ablation}.

\subsection{Online Evaluation via Multi-aspect Online Metrics (RQ2)}
\label{sec.online_evaluation}

\subsubsection{Evaluation Protocols.}

We deploy MicRec on a widely-used recommender system as Sec. \ref{sec.online_deployment} to show the power of MicRec in online.
The online baseline is ICAN with other online modules unchanged.
We deploy MicRec for the video domain. For the overall system, we concentrate on the following four online metrics: (1) average click count per capita (ACC), (2) average dwell time per capita (ADT), (3) average click count of cold-start users (ACC-c), and (4) average dwell time of cold-start users (ADT-c). For metrics on videos, we further measure: (5) average like rate per capita (ALR), and (6) average follow rate per capita (AFR) besides ACC and ADT.
We conduct the online A/B test for $7$ days on $4$ million users.

\begin{table}[!hbtp]
\centering
\small
\caption{Online A/B tests. All improvements are significant (t-test with $p<0.05$).}
\begin{tabular}{p{1.6cm}|p{1.6cm}<{\centering}p{1.6cm}<{\centering}p{1.6cm}<{\centering}p{1.6cm}<{\centering}}
\toprule
Overall & ACC & ADT & ACC-c & ADT-c \\
\midrule
MicRec & +3.287\% & +1.389\% & +11.517\% & +6.472\%  \\
\bottomrule
\end{tabular}
\begin{tabular}{p{1.6cm}|p{1.6cm}<{\centering}p{1.6cm}<{\centering}p{1.6cm}<{\centering}p{1.6cm}<{\centering}}
\toprule
Video & ACC & ADT & ALR & AFR \\
\midrule
MicRec & +5.643\% & +2.824\% & +4.138\% & +5.165\%  \\
\bottomrule
\end{tabular}
\label{tab:online}
\end{table}

\subsubsection{Experimental Results.}

Table \ref{tab:online} displays the improvement percentages of MicRec over the online base matching model, from which we can find that:

(1) MicRec achieves significant improvements on all online metrics (significance level $p<0.05$), which confirms the effectiveness of MicRec in online systems. ACC focuses on the central click-related metrics. ADT focuses on the user duration, which can reduce the impact of clickbaits and reflect users' real satisfaction. The consistent improvements imply that MicRec could retrieve more high-quality items to be recommended to users in matching.

(2) From ACC-c and ADT-c, we know that cold-start users get more improvements from MicRec. It is because that our three CL tasks can introduce additional item information to benefit representation learning, which indirectly enhances the predictions of cold-start users. We also notice that there is an astonishing $1.54\%$ improvement on the next-day user retention rate of cold-start users, which implies that MicRec even improves the user stickiness to the system.

(3) In the video scene, besides the consistent improvements of ACC and ADT, we also observe that MicRec outperforms the base model on ALR and AFR. Our system encourages users' diverse active interactions such as \emph{like} and \emph{follow}, and the improvements on like- and follow- related metrics reconfirm users' satisfaction with our recommended items on diverse online metrics.

\subsection{Ablation Study on Accuracy and Diversity (RQ3)}
\label{sec.ablation}

To verify the effectiveness of three item-based CL tasks of MicRec, we conduct four ablation versions without certain CL tasks (\emph{fea}, \emph{sem}, \emph{sess}) on Video-636M.
Table \ref{tab:ablation_test} shows the HIT@N and coverage results of different versions, we have:

\begin{table}[!hbtp]
\centering
\small
\caption{Impacts of three CL tasks on accuracy (HIT@N) and diversity (coverage).}
\begin{tabular}{l|p{2.0cm}<{\centering}p{2.0cm}<{\centering}p{2.0cm}<{\centering}}
\toprule
Ablation version & HIT@50 & HIT@500 & item coverage \\
\midrule
MicRec & 0.2047 & 0.5209 & 16.1\%\\
\midrule
  \quad -- Feature-level item CL & 0.2008 & 0.5092 & 17.2\%\\
  \quad -- Semantic-level item CL & 0.2033 & 0.5134 & 16.2\% \\
  \quad -- Session-level item CL & 0.2012 & 0.5108 & 9.3\% \\
  \quad -- all item-based CL & 0.1872 & 0.4873 & 12.4\% \\
\bottomrule
\end{tabular}
\label{tab:ablation_test}
\end{table}

(1) MicRec performs worse when any item-based CL loss is removed, which reconfirms the power of all multi-granularity CL tasks. Note that MicRec's improvements are significant ($p<0.05$ over MicRec's ablation versions and $p<0.01$ over other baselines). It implies that MicRec can alleviate the sparsity issue via SSL with feature, semantic, and session based item augmentations.

(2) The feature-level item CL conducts feature dropouts to build fine-grained feature-level augmentations for items. It is an effective and classical way to introduce additional SSL signals for sufficient training, improving the \emph{alignment} between items and their augmentations and the \emph{uniformity} of all item representations, and thus enable better performances in matching.

(3) The semantic-level item CL focuses on the coarse-grained semantic similarities between two item titles. We highlight this semantic information since it is directly exposed to users and is one of the main factors largely impacting users. This CL also brings in significant improvements besides other two CL tasks.

(4) The session-level item CL highlights the global behavioral correlations between items according to the ``unconscious crowd-sourcing'' by users' natural clicks in sessions. The session information is an essential supplement to the user-item click behaviors in discovering users' sequential behavioral preferences.

(5) From coverage, we find that the session-level item CL is the major source of diversity improvements. The diverse and less predictable sequential behaviors accumulated from the global sessions can bring in more high-quality occasionality for breaking the filter bubble and exploring more possibilities. Its huge diversity improvement (from $9.3\%$ to $16.1\%$) also implies that items detected by session-level item correlations cannot be easily covered by ICAN and the other two CL tasks.
In contrast, models with feature-level and semantic-level CL tasks have comparable or even slightly worse diversity results (which might be solved if we conduct in-batch negative sampling as \cite{zhou2021contrastive}, while it will sacrifice the accuracy in our systems). However, the diversity metric is only an indirect metric, not the final purpose (i.e., improving user satisfaction). Therefore, we still jointly conduct all three item-based CL tasks in our final MicRec, since it has the best accuracy performance with relatively good diversity performance compared to other versions. Such a setting is also verified by real users as stated in Sec. \ref{sec.online_evaluation}.

\subsection{Universality of MicRec on Different Matching Models (RQ4)}
\label{sec.universality}

To demonstrate MicRec's universality, we further deploy our item-based CL tasks on other matching models. Fig. \ref{fig.universality} shows the HIT@N results of different proposed CL combinations on YoutubeDNN \cite{covington2016deep} and AutoInt \cite{song2019autoint}. We can observe that:

\begin{figure*}[!htbp]
\centering
\subfigure[Results on HIT@50.]{
\label{Fig.sub.1}
\includegraphics[width=0.30\textwidth]{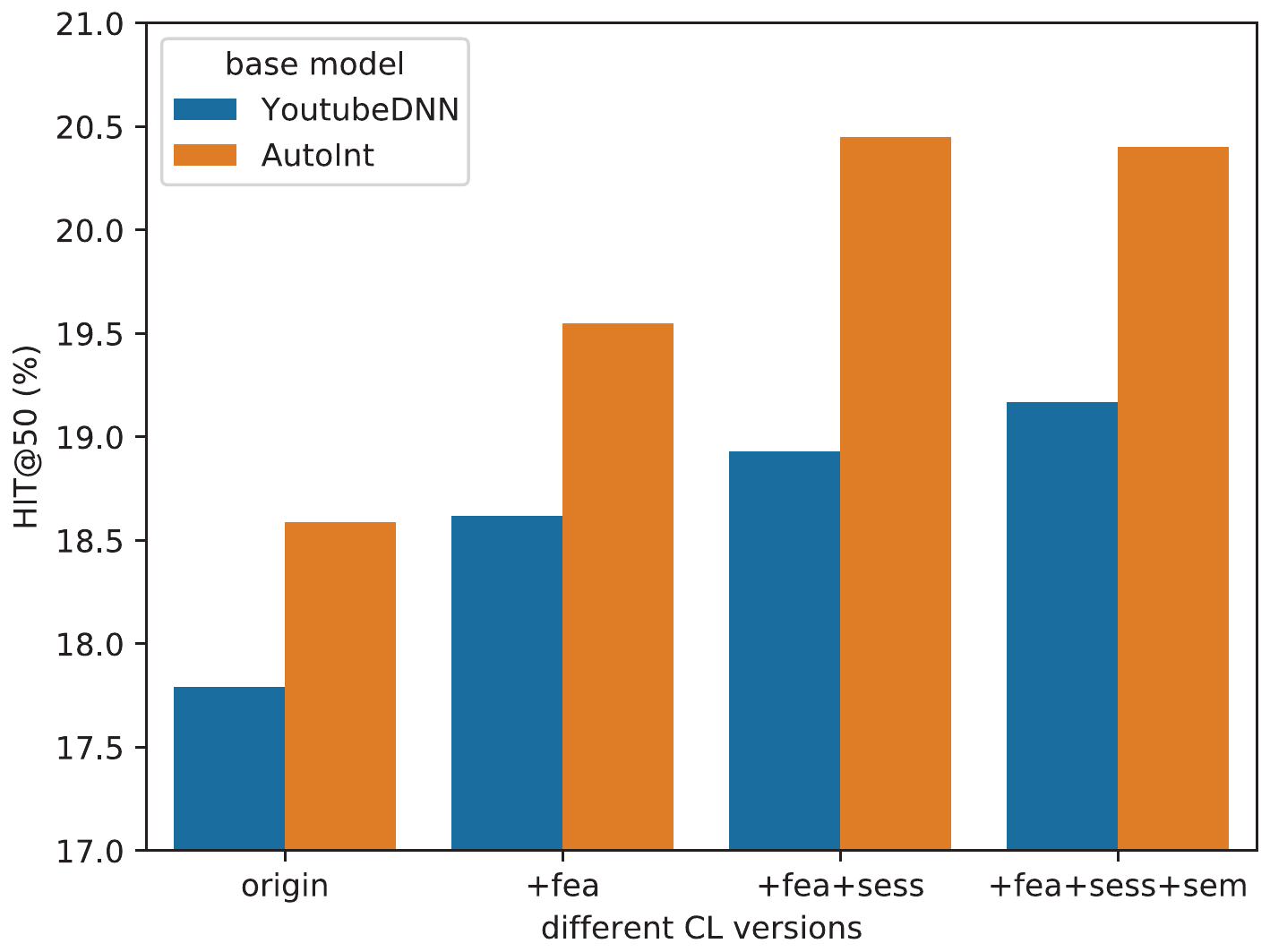}}
\subfigure[Results on HIT@200.]{
\label{Fig.sub.2}
\includegraphics[width=0.30\textwidth]{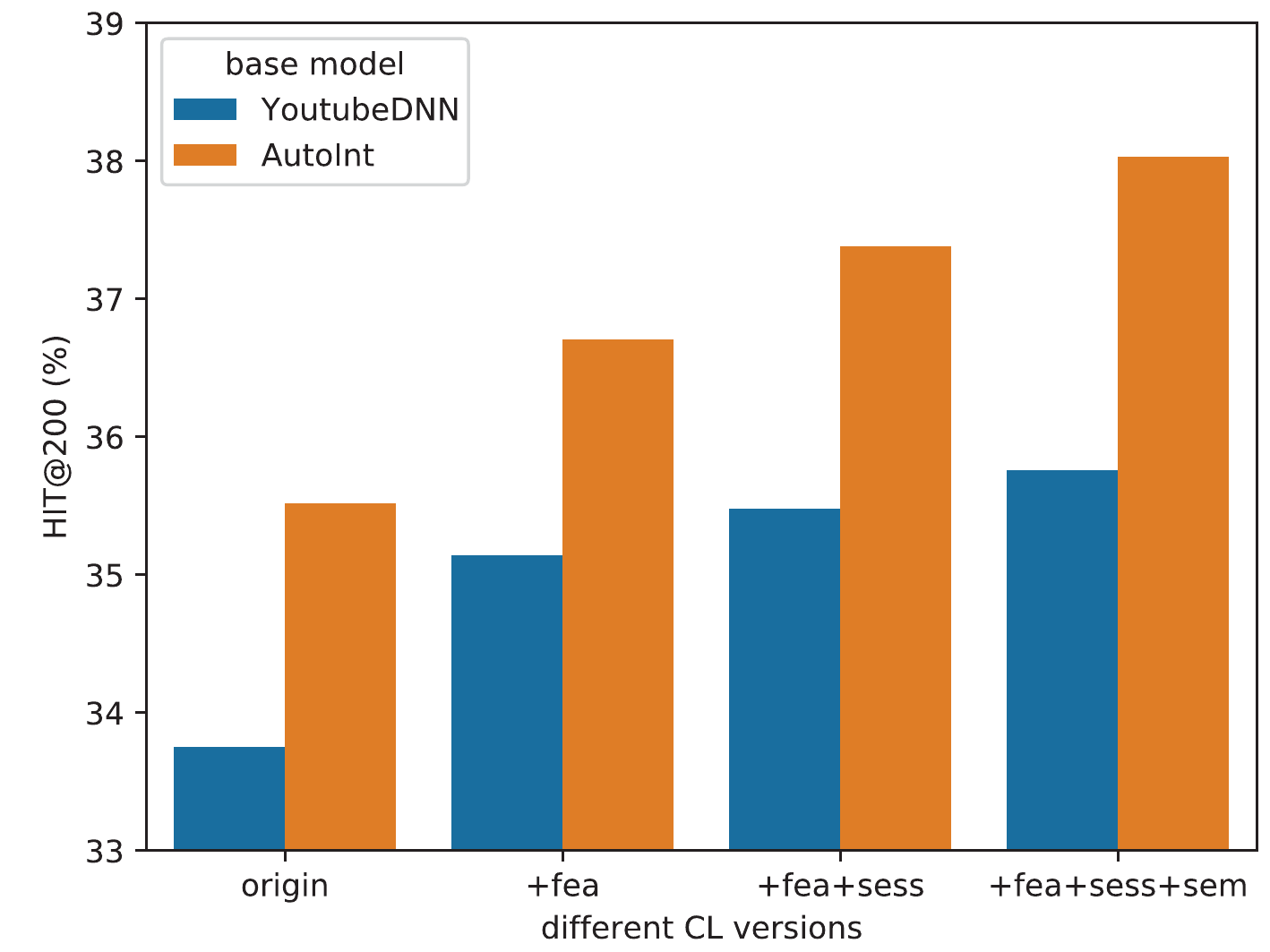}}
\subfigure[Results on HIT@500.]{
\label{Fig.sub.3}
\includegraphics[width=0.30\textwidth]{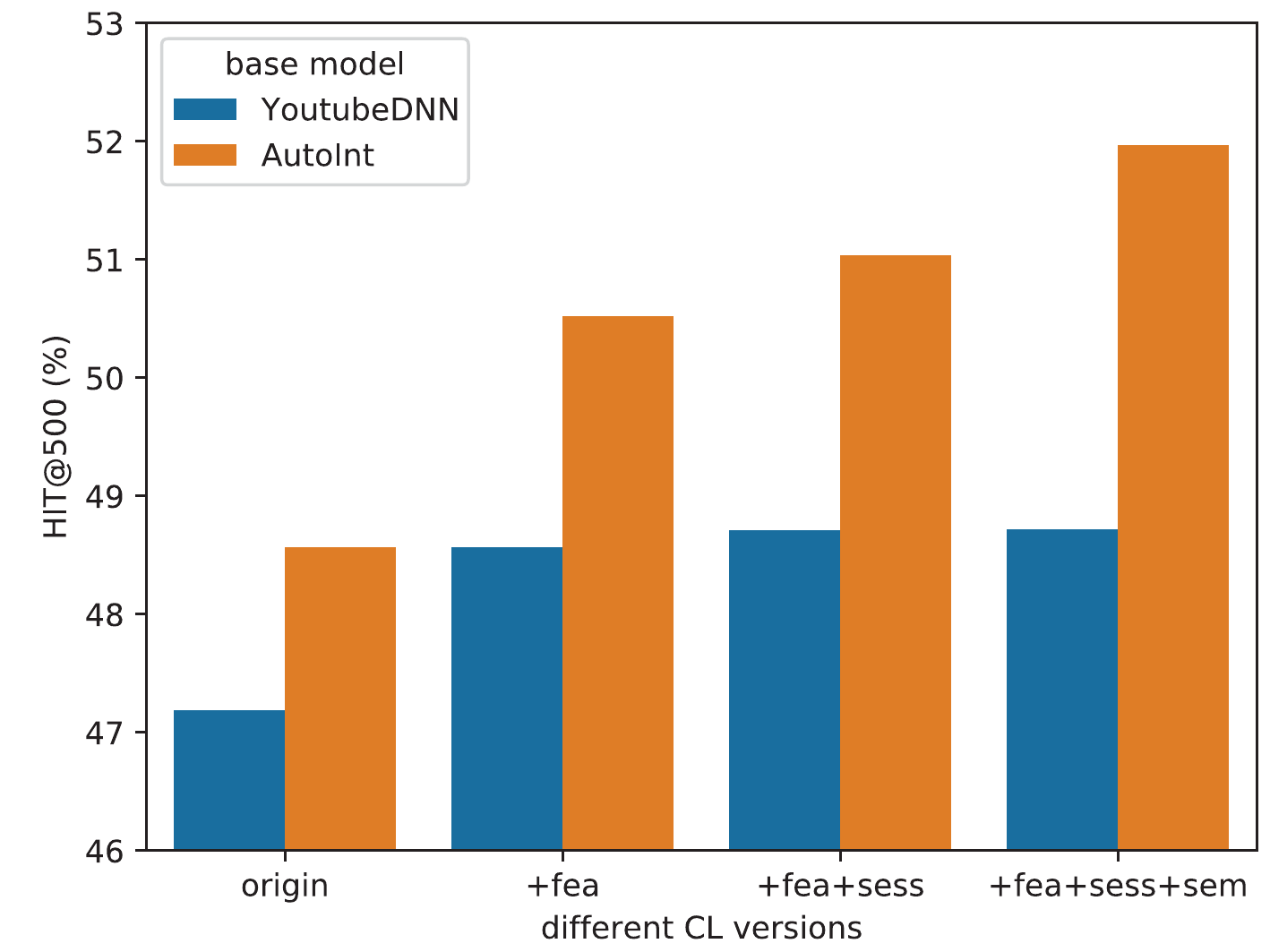}}
\caption{Improvements of three CL tasks with two matching models on Video-636M.}
\label{fig.universality}
\end{figure*}

(1) In general, all feature-level, semantic-level, and session-level item CL tasks are effective even deployed with other models (while MicRec still has the best performances). It demonstrates the effectiveness and universality of MicRec and all proposed item-based CL tasks with different matching models.

(2) The feature-level CL and session-level CL tasks contribute the main part of the significant improvements as the conclusion in Sec. \ref{sec.ablation}. To emphasize the universality of three CL tasks, we directly follow the hyper-parameters of MicRec with ICAN, which already achieves satisfactory improvements. MicRec is not that dependent on parameter engineering, which is developer-friendly in practice.

\section{Conclusions and Future Work}

In this work, we propose a novel MicRec framework, which designs three effective and universal feature-level, semantic-level, and session-level item-based CL to improve the matching performances. In real-world systems, MicRec achieves significant online and offline improvements with different matching models and datasets, and has been deployed in online.
Although the proposed three item-based CL tasks seem to be straightforward, we have verified their effectiveness and universality on different scenarios and models in offline and online, highlighting the under-explored but promising directions of item-based CL in practice.

In the future, we will design more effective item-based CL tasks, and jointly combine user-based CL with item-based CL tasks. We will also explore the universality of MicRec in ranking and better CL-based optimization strategies.

 \bibliographystyle{splncs04}
 \bibliography{reference}

\end{document}